\documentclass[aps,pra,amsmath,amssymbs,superscriptaddress]{revtex4}

\usepackage{graphicx}
\usepackage{amsmath,amsfonts,amssymb}
\usepackage{epsfig}
\usepackage{wrapfig}
\usepackage{bbm}
\usepackage[usenames]{color}
\usepackage{array}
\usepackage{times}
\usepackage{setspace}
\usepackage{epstopdf}

\begin{document}
%%%%%%%%%%%%%%%%%%%%%%%%%%%%%%%%%%%%%%%%%%%%%%%%%%%%%%%%%%%%%%%%%%%%%%%%%%%%%%%%%%%%%%%%%%%%%%%%%%%%%%%%%%%%%%%%%%%%%%%%%%%%
% Paper Header
%%%%%%%%%%%%%%%%%%%%%%%%%%%%%%%%%%%%%%%%%%%%%%%%%%%%%%%%%%%%%%%%%%%%%%%%%%%%%%%%%%%%%%%%%%%%%%%%%%%%%%%%%%%%%%%%%%%%%%%%%%%%

\title{Measurement of Topological Invariants in a 2D Photonic System}
\author{
Sunil Mittal$^{1,2}$, Sriram Ganeshan$^{1,3}$, Jingyun Fan$^{1}$, Abolhassan Vaezi$^{4}$, \\
Mohammad Hafezi$^{1,2,\dagger}$
\\
\normalsize{${}^{1}$Joint Quantum Institute, NIST/University of Maryland, College Park MD 20742, USA} \\
\normalsize{${}^{2}$Department of Electrical Engineering and Institute for Research in Electronics and} \\
\normalsize{Applied Physics, University of Maryland, College Park MD 20742, USA}\\
\normalsize{${}^{3}$Condensed Matter Theory Center, University of Maryland, College Park MD 20742, USA}\\
\normalsize{${}^{4}$Laboratory of Atomic and Solid State Physics, Cornell University, Ithaca NY 14853, USA}\\
\normalsize{$\dagger$\textit{hafezi@umd.edu}}}

\maketitle
\setstretch{2}

%%%%%%%%%%%%%%%%%%%%%%%%%%%%%%%%%%%%%%%%%%%%%%%%%%%%%%%%%%%%%%%%%%%%%%%%%%%%%%%%%%%%%%%%%%%%%%%%%%%%%%%%%%%%%%%%%%%%%%%%%%%%
% Abstract
%%%%%%%%%%%%%%%%%%%%%%%%%%%%%%%%%%%%%%%%%%%%%%%%%%%%%%%%%%%%%%%%%%%%%%%%%%%%%%%%%%%%%%%%%%%%%%%%%%%%%%%%%%%%%%%%%%%%%%%%%%%%

\textbf{A hallmark feature of topological physics is the presence of one-way propagating chiral modes at the system boundary \cite{wenbook,bernevigbook}. The chirality of edge modes is a consequence of topological character of the bulk. For example, in a non-interacting quantum Hall (QH) model, edge modes manifest as mid-gap states between two topologically distinct bulk bands. The bulk-boundary correspondence dictates that the number of chiral edge modes, a topological invariant called the winding number, is completely determined by the bulk topological invariant, the Chern number \cite{Hatsugai1993}. Here, for the first time, we measure the winding number in a two-dimensional (2D) photonic system. By inserting a unit flux quantum to the edge, we show that the edge spectrum resonances shift by the winding number. This experiment provides a new approach for unambiguous measurement of topological invariants, independent of the microscopic details, and could possibly be extended to probe strongly correlated topological orders.}

%%%%%%%%%%%%%%%%%%%%%%%%%%%%%%%%%%%%%%%%%%%%%%%%%%%%%%%%%%%%%%%%%%%%%%%%%%%%%%%%%%%%%%%%%%%%%%%%%%%%%%%%%%%%%%%%%%%%%%%%%%%%
% Main Text
%%%%%%%%%%%%%%%%%%%%%%%%%%%%%%%%%%%%%%%%%%%%%%%%%%%%%%%%%%%%%%%%%%%%%%%%%%%%%%%%%%%%%%%%%%%%%%%%%%%%%%%%%%%%%%%%%%%%%%%%%%%%

Recently, there has been a surge of interest in investigating topological states with synthetic gauge fields. Synthetic gauge fields have been realized in various atomic \cite{Miyake2013, Jotzu2014, Spielman2013, Aidelsburger2015} and photonic systems \cite{Hafezi2014PT, Ling2014}. In particular, topological photonic edge states have been imaged in two recent concurrent experiments \cite{Hafezi2013, Rechtsman2013} and robustness of their transport has been quantitatively confirmed both in the microwave \cite{Wang2009} and telecom domains \cite{Mittal2014}. Several other interesting proposals have investigated topological states in 1D \cite{Kraus2012, Hu2014, Zeuner2014}, 2D \cite{Umucalilar2011,Fang2012,Khanikaev2013,Tzuang:2014,Ma2015} and also 3D \cite{Lu:2015} synthetic structures. Topological states are characterized by topologically invariant integers \cite{bernevigbook}. More specifically,  the bulk-boundary correspondence dictates that the number of chiral edge modes, the winding number, is completely determined by the bulk topological invariant, the Chern number \cite{Hatsugai1993}. In fermionic systems, conductance measurements reveal these quantum numbers. However, a direct measurement of these integers is non-trivial in bosonic systems, mainly because the concept of conductance is not well defined \cite{Ozawa2014, Hafezi2014}. While these integers have been measured in 1D bosonic systems \cite{Atala2013,Hu2014,Duca2015}, the 2D bosonic case has been limited to atomics lattices \cite{Aidelsburger2015}.

Here, we experimentally demonstrate that selective manipulation of the edge can be exploited to measure topological invariants, i.e., the winding number of the edge states. We implement an integer quantum Hall system using a fixed, uniform synthetic gauge field in the bulk and couple an additional tunable gauge field only to the edge. The edge state energy spectra flows as a function of this tunable flux. With insertion of a unit quantum of flux, the edge state resonances move by $\pm 1$, which is the winding number of edge states in our system. This spectral flow can be directly observed in an experiment as the flow of transmission resonances, and thus provides a direct measurement of the winding number of edge states. For this demonstration, we employ the unique ability of our photonic system to selectively manipulate edge states --- a feature that is challenging to achieve in current electronic and atomic systems.

To model the spectral flow of a quantum Hall edge, with winding number $k=1$, we consider a linear edge dispersion $E_p = v p$ where $E_p$ is the energy, $v$ is the group velocity, and $p$ is the momentum along the edge.  When a gauge flux ($\theta$) is coupled to the edge, the momentum is replaced by the covariant momentum, i.e.,
\begin{equation}
E_p=v\left(p-q\frac{\theta}{L}\right),
\end{equation}
where $L$ is the length of the edge and $q$ is the charge of the edge excitations. For non-interacting photons, the charge $q=1$. Note that the corresponding vector potential is simply $\theta/L$. For a finite system, quantization of momentum on the edge results in
\begin{equation}
E_n= \frac{2\pi v}{L} \left(n-\frac{\theta}{2\pi}\right),
\end{equation}
where $n$ is an integer. Thus, insertion of $\theta=2\pi$ flux shifts $E_n\to E_{n-1}$  resulting in an {\it anomalous spectral flow}, as shown in Fig. 1a. This is in contrast to the case of a topologically trivial system, where there is no net shift and having a zero and $2\pi$ fluxes are equivalent to each other. However, in QH systems, these linear-dispersion mid-gap modes continuously interpolate between two topologically distinct bulk bands and the difference between the Chern numbers of these bulk bands dictates the shift, and hence the winding number of the edge states. For winding numbers larger than one  $(k > 1)$, there exist $k$ edge modes in the bandgap and hence $k$ copies of this relation, i.e., for a unit quantum of flux insertion, the edge state resonances shift by $k$ units. Such spectral flow is similar to Laughlin's charge pump \cite{Laughlin1981, Halperin1982} but with an important distinction: here, the gauge flux is coupled only to the edge of the system (see SI). Moreover, the observable in our interferometric measurement is expected to be an integer, in contrast to the scheme presented in \cite{Ozawa2014} which measures Hall drift, a continuous variable. More generally, our scheme provides a powerful universal probe (independent of microscopic details) of topological order that can be generalized to the case of strongly correlated topological systems.

\linespread{1.0}
\begin{figure}
 \centering
 \includegraphics[width=0.98\textwidth]{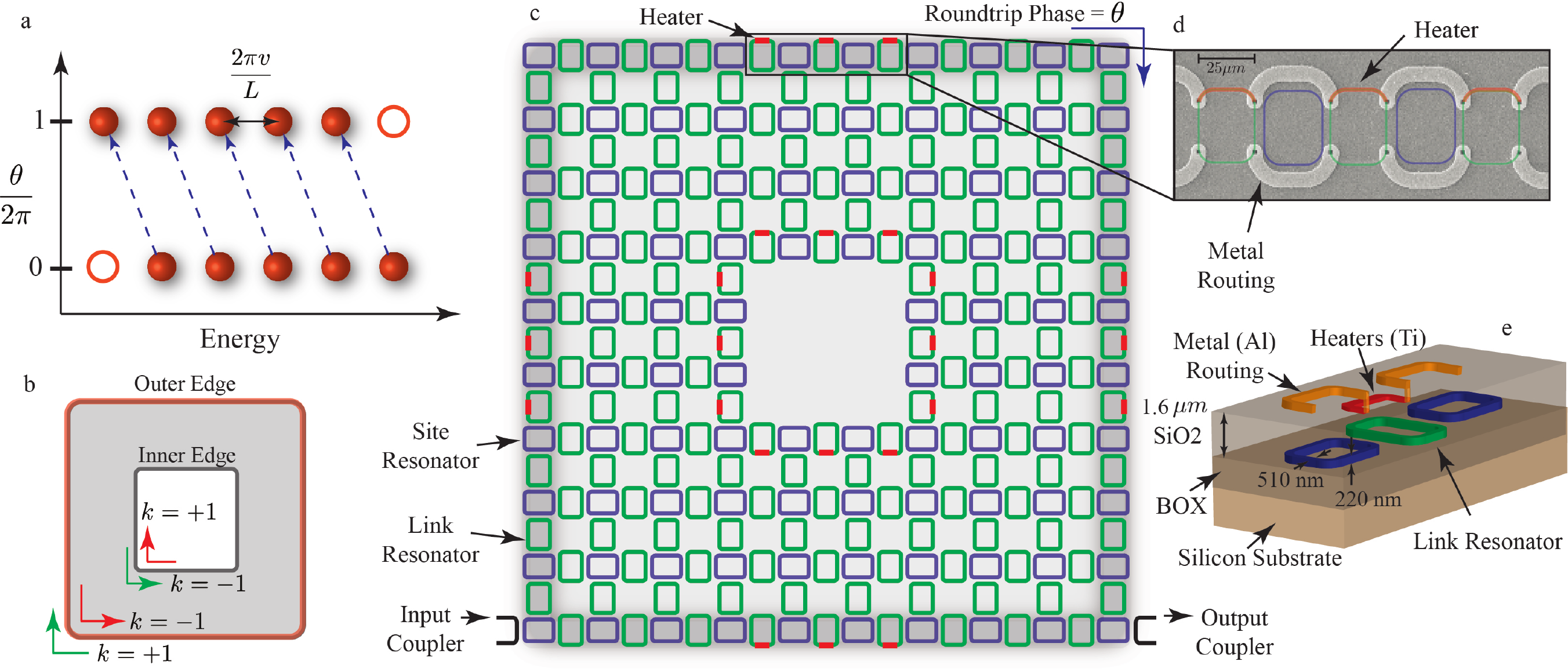}
 \caption{\textbf{a}, Spectral flow of edge states with coupled gauge flux $\theta$ for $q=1$ and $k=1$.  When $\theta=2\pi$, edge states move by $k=1$ positions. \textbf{b,c}, Square annulus of coupled ring resonators which implements the integer quantum Hall effect, with outer edge of 10 sites and inner edge of 4 sites. The uniform synthetic magnetic field for photons is introduced by asymmetrically positioning the link rings. The system exhibits two pairs of clockwise and counter clockwise propagating edge states, one each on the outer and the inner edges. The tunable gauge field coupled only to the edge states is introduced by fabricating heaters on link resonators situated on the lattice edges. The total flux $\theta$ introduced by heaters is the sum of individual phases incurred in link resonator arms. \textbf{d}, SEM image showing heaters fabricated on top of link resonators (green). \textbf{e}, Schematic of the waveguide cross section showing the ring resonators, the metal heaters and the metal routing layer.}
 \label{fig:1}
\end{figure}
\linespread{2.0}

To experimentally observe and measure this spectral flow, we implement the integer quantum Hall model in a photonic system: a two dimensional square annulus of ring resonators with a uniform synthetic magnetic field in the bulk and a tunable gauge field coupled only to the lattice edge (Fig. 1b-e). The uniform magnetic field with flux $\phi_{0} = \frac{2\pi}{4}$ radians per plaquette is synthesized using asymmetric placement of site and link resonators, as previously described in Ref.\cite{Hafezi2013}. To couple a tunable gauge field to the edges, we fabricate metal heaters above the link ring waveguides on the lattice edge (Fig. 1c-e). These heaters use thermo-optic effect to modify the accumulated phase of light propagating through the waveguides and hence result in a gauge flux.

\linespread{1.0}
\begin{figure}
 \centering
 \includegraphics[width=0.98\textwidth]{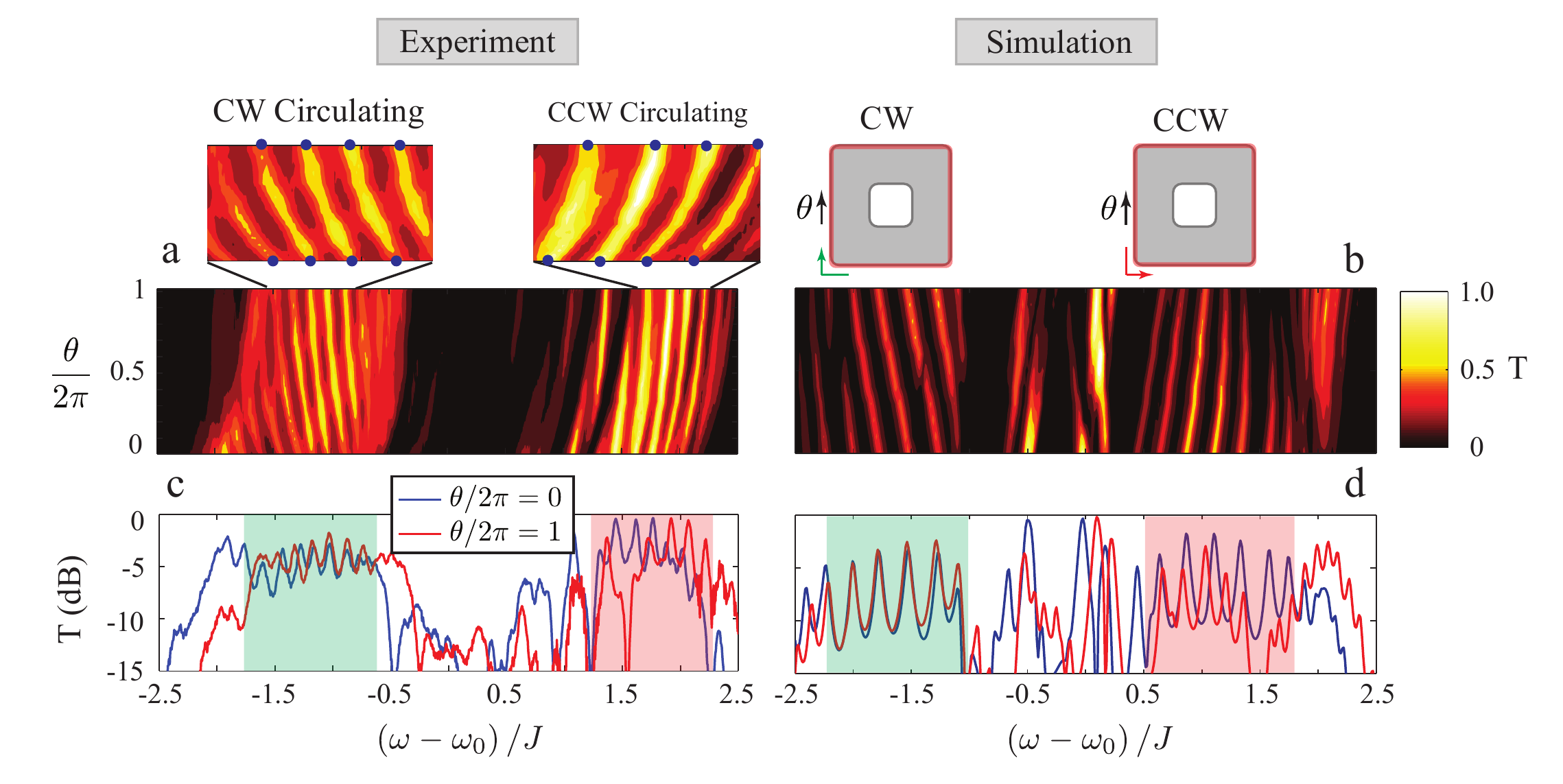}
 \caption{\textbf{a}, Measured and \textbf{b}, simulated transmission ($T$, in linear scale) as a function of phase $\theta$ and $\omega$ (frequency), for spin-up excitation. For a $2\pi$ increase in gauge flux, the edge state resonances move by one position, giving $k = +1.0 (1)$ for CW edge band and $k = -1.0 (2)$ for the CCW band. Insets: Zoom-in of the edge state bands. \textbf{c}, Measured and \textbf{d}, simulated transmission spectra (log scale) for $\theta/2\pi = 0, 1$. For this $2\pi$ increase in flux, the measured spectra match approximately in the edge state regions. The green and red shaded regions indicate the
 CW and CCW edge state bands, respectively. The transmission is normalized such that the maximum is unity. Here $\omega_{0}$ is the resonance frequency of the resonators and $J$ is the coupling rate between site resonators.}
 \label{fig:2}
\end{figure}
\linespread{2.0}

Our system supports two pseudo-spin components, up and down, which circulate in opposite directions in ring resonators \cite{Hafezi2013}. Each pseudo-spin component experiences a uniform magnetic field, and we can selectively excite each pseudo-spin by appropriately choosing the input port (Fig. 1c). Furthermore, with this particular choice of uniform magnetic field, the energy spectrum exhibits two bandgap regions \cite{Hatsugai1993}. For spin-up excitation, the low energy bandgap corresponds to clockwise (CW) circulating edge states confined on the outer edge,  and counter-clockwise (CCW) circulating edge states on the inner edge, as shown with green arrows in Fig. 1b with their corresponding winding numbers. Similarly, the high energy bandgap corresponds to edge states both on the outer and the inner edges, however, they circulate opposite to those in the low energy bandgap, as shown in Fig. 1b by red arrows. A pseudo-spin flip swaps the position of CW and CCW edge bands in the transmission spectrum, and also flips the sign of gauge flux $\theta$ \cite{Hafezi2011}.

Fig. 2a,b show the measured and simulated transmission spectra as a function of the coupled flux $\theta$, for spin-up excitation. Edge states of the outer edge (bright regions) and the bulk states (dark region in the middle) are easily identifiable. As the coupled flux $\theta$ increases, the energy of the CW edge states decreases, whereas the energy of CCW edge states increases. For a $2\pi$ increase in  flux, the edge state resonances move by one resonance to replace the position once held by its neighbor. This flow indicates that the measured winding number is $k = +1.0 (1)$ for the CW circulating edge states, and $k=-1.0 (2)$ for the CCW circulating edge states. We note that only the outer edge contributes to the transmission spectra, because the edge states are tightly confined to the lattice boundary. Moreover, such sharpness of edge states allows us to locally create the extra gauge flux $\theta$ by placing the heaters only on the outmost rings at the boundary. Also note that the flow of resonances observed here is very distinct in appearance and physical origin from what is observed in a Fabry-Perot cavity, where increasing the length of the cavity always reduces the resonant frequencies. However, in our observation CW and CCW edge state resonances shift in opposite directions, and resonance peaks shift exactly by one.

Fig. 2c,d show the overlap of the observed and simulated transmission spectra at $\frac{\theta}{2\pi} = 0$ and $1$, showing excellent qualitative agreement with the simulation results. We attribute the small discrepancy in the overlap to lattice disorder introduced by the nonlocal response of the heaters (see SI). The discrepancy is more pronounced for the CCW state because of finite size effects in travelling a shorter path from input to the output port. Also note that compared to measured spectra, simulated spectra show high transmission through the two bulk bands. This is because the bulk bands are not topologically protected and therefore, disorder in the experimental system suppresses transmission through these bands.

\linespread{1.0}
\begin{figure}
 \centering
 \includegraphics[width=0.98\textwidth]{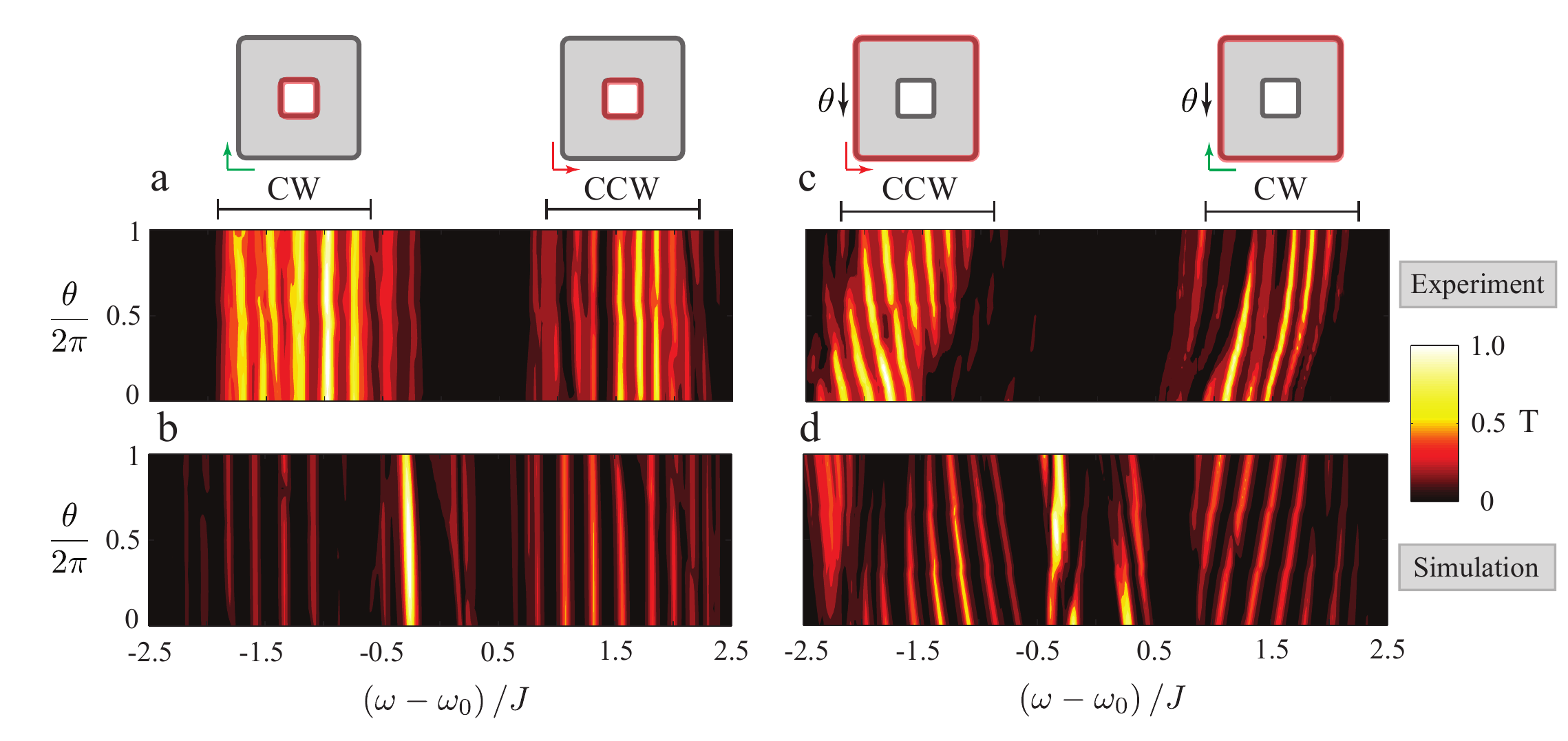}
 \caption{\textbf{a}, Measured and \textbf{b}, simulated transmission spectrum (linear scale, normalized to unity)  as a function of  flux $\theta$ when only the inner edge is heated. The energies of the edge states associated with the outer edge do not change. \textbf{c}, Measured and \textbf{d} simulated transmission spectrum, for spin-down excitation. The CW and CCW circulating states have now exchanged positions and move in opposite directions (compared to Fig. 2a,b), giving $k = 1.2 (2)$ for CW edge band and $k = -1.2 (2)$ for the CCW band. The transmission spectra shown in \textbf{a} and \textbf{c} are from two different devices.}
 \label{fig:3}
\end{figure}
\linespread{2.0}

In order to verify the local character of the gauge coupling, we selectively couple a nonzero flux $\theta$ to the inner edge by heating the link rings on the inner edge. We observe that heating only the inner edge does not shift the edge states (Fig. 3a,b), which indicates the strong confinement of edge states at the lattice boundary. Furthermore, we excited the system with a flipped pseudo-spin (compared to Fig. 2). Fig. 3c,d shows the measured and simulated transmission spectra as a function of coupled flux $\theta$. Since the spin flip reverses both the sign of the coupled flux $\theta$ and the position of edge bands in the transmission spectrum, the resulting spectrum is similar to Fig.2.

\linespread{1.0}
\begin{figure}
 \centering
 \includegraphics[width=0.98\textwidth]{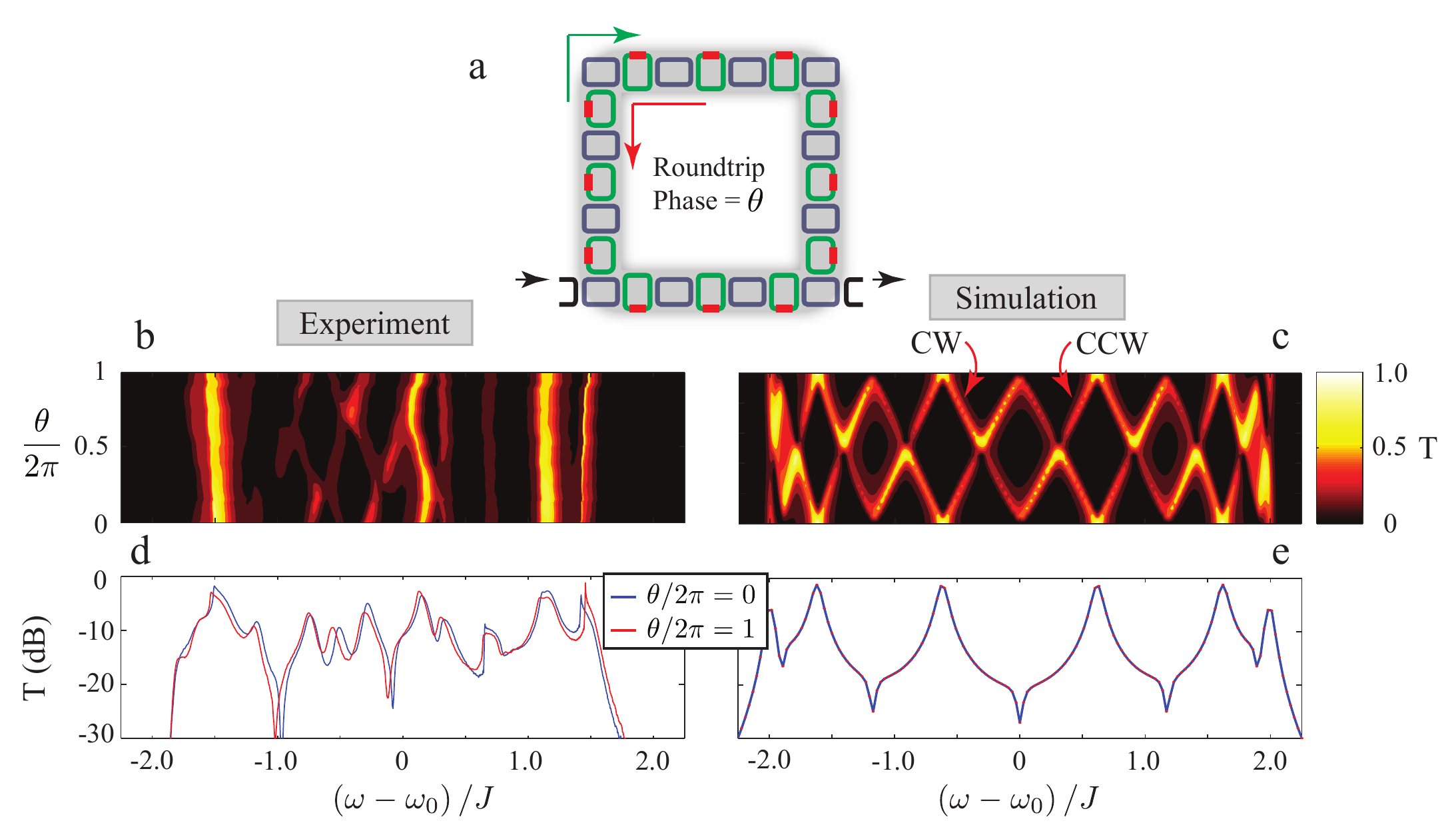}
 \caption{\textbf{a} A ring geometry to investigate gauge coupling to topologically trivial states. The ring supports CW and CCW circulating modes. External flux can be coupled to these states by using heaters on the link resonators. \textbf{b}, Measured and textbf{c}, simulated transmission as a function of external flux $\theta$ and frequency $\omega$. The CW and CCW modes move in opposite directions but for a unit increase in flux, they do not change position. \textbf{d}, Measured and \textbf{e}, simulated transmission spectra at $\theta = 0, 2\pi$ overlap well.}
 \label{fig:4}
\end{figure}
\linespread{2.0}

As a control experiment, to investigate the difference of gauge coupling to chiral and non-chiral states, we fabricate a ring of resonators (Fig. 4a) which corresponds to a 1D tight-binding model. This system supports topologically trivial Bloch modes, circulating in CW and CCW directions, which are degenerate in the absence of a magnetic flux. Fig. 4b,c show the experimentally observed and the simulated transmission spectra as a function of coupled flux $\theta$. The flux lifts the degeneracy of the CW and CCW states and with an increase in the flux, these states move in opposite directions. Therefore, for $\theta=2\pi$, there is no net shift of the resonances, with measured $k=0.03(8)$. The absence of spectral flow proves that this system is topologically trivial. As shown in Fig. 4d, we observe the transmission spectra at $\theta/2\pi = 0, 1$ exhibit an excellent overlap. However, compared to the simulation, we realize that not all the states can be resolved in the experimental data. Such discrepancy is due to the absence of topological protection of non-chiral states against disorder which leads to coupling between CW and CCW modes.

Here, we used a simple model to describe spectral flow of a chiral edge coupled to a gauge field, for non-interacting bosons. However, the emergence of anomalous spectral flow is more general \cite{wenbook,bernevigbook}, and can be modeled using the elegant framework of conformal field theory of chiral bosons, where this behavior is known as chiral anomaly \cite{wenbook,ivan09, sriramthesis}. This framework allows the investigation of systems with strong photon-photon interaction, such as circuit-QED \cite{Houck2012} and polariton \cite{Carusotto2013} systems, and also ultracold atoms with topological features \cite{Atala2013, Miyake2013, Jotzu2014,Spielman2013}. In particular, the presence of strong interactions could lead to the emergence of various fractional quantum states in the steady-state regime \cite{Kapit2014}. Spectral flow could be an interesting way to explore the fascinating many-body features of topological states in systems with synthetic gauge fields.

\textbf{Methods}

\textbf{Heaters:}
The heaters used in these devices are 110 nm thick metal (Ti) pads, with typical resistance of 117 Ohms, such that a current flowing through them generates heat and modifies the refractive index, and the accumulated phase of light propagating through the waveguide. A typical heater requires $\approx$ 76.5 mW to introduce a $2\pi$ phase. Instead of employing a single heater which may introduce a significant disorder in the lattice, we distribute the heaters along the edges while maintaining an accumulated flux of $2\pi$ (see SI).

\textbf{Measurement of winding number:}
To estimate the value of winding number $k$, we measure the frequency shift of an edge state resonance for $2\pi$ flux insertion; and its separation from the neighboring resonance at $\theta = 0$. The integer $k$, in a given band, is then the mean of the ratio of shift of resonances to their separation. The standard deviation of this ratio is the error on $k$.

%%%%%%%%%%%%%%%%%%%%%%%%%%%%%%%%%%%%%%%%%%%%%%%%%%%%%%%%%%%%%%%%%%%%%%%%%%%%%%%%%%%%%%%%%%%%%%%%%%%%%%%%%%%%%%%%%%%%%%%%%%%%
% Bibliography
%%%%%%%%%%%%%%%%%%%%%%%%%%%%%%%%%%%%%%%%%%%%%%%%%%%%%%%%%%%%%%%%%%%%%%%%%%%%%%%%%%%%%%%%%%%%%%%%%%%%%%%%%%%%%%%%%%%%%%%%%%%%

\bibliographystyle{NatureMag}
\bibliography{Winding_Biblio}

%%%%%%%%%%%%%%%%%%%%%%%%%%%%%%%%%%%%%%%%%%%%%%%%%%%%%%%%%%%%%%%%%%%%%%%%%%%%%%%%%%%%%%%%%%%%%%%%%%%%%%%%%%%%%%%%%%%%%%%%%%%%
% Acknowledgements
%%%%%%%%%%%%%%%%%%%%%%%%%%%%%%%%%%%%%%%%%%%%%%%%%%%%%%%%%%%%%%%%%%%%%%%%%%%%%%%%%%%%%%%%%%%%%%%%%%%%%%%%%%%%%%%%%%%%%%%%%%%%

\textbf{Acknowledgements}

This research was supported by Air Force Office of Scientific Research grant no. FA9550-14-1-0267, Army Research Office - Multidisciplinary University Research Initiative, Bethe postdoctoral fellowship, National Science Foundation Career grant, Laboratory for Physical Sciences -Condensed Matter Theory Center, Microsoft and the Physics Frontier Center at the Joint Quantum Institute. We thank M. Levin, A. G. Abanov, A. Lobos, A. Migdall, J. Taylor for fruitful discussions and E. Barnes, M. Davanco, E. Goldschmidt for useful comments on the manuscript.

\textbf{Author contributions}

S.M. and M.H. conceived and designed the experiment. S.M. and J.F. performed the experiment. All authors contributed significantly in analyzing the data and editing the manuscript.

\textbf{Additional information}

Supplementary information is available in the online version of the paper. Reprints and permissions information is available online at www.nature.com/reprints. Correspondence and requests for materials should be addressed to M.H.

\textbf{Competing financial interests}

Authors declare no competing financial interests.

\end{document}

\title{Supplementary Information: Measurement of Topological Invariants in a 2D Photonic System}
\author{
Sunil Mittal$^{1,2}$, Sriram Ganeshan$^{1,3}$, Jingyun Fan$^{1}$, Abolhassan Vaezi$^{4}$, \\
Mohammad Hafezi$^{1,2,\dagger}$
\\
\normalsize{${}^{1}$Joint Quantum Institute, NIST/University of Maryland, College Park MD 20742, USA} \\
\normalsize{${}^{2}$Department of Electrical Engineering and Institute for Research in Electronics and} \\
\normalsize{Applied Physics, University of Maryland, College Park MD 20742, USA}\\
\normalsize{${}^{3}$Condensed Matter Theory Center, University of Maryland, College Park MD 20742, USA}\\
\normalsize{${}^{4}$Laboratory of Atomic and Solid State Physics, Cornell University, Ithaca NY 14853, USA}\\
\normalsize{$\dagger$\textit{hafezi@umd.edu}}}

\maketitle

%%%%%%%%%%%%%%%%%%%%%%%%%%%%%%%%%%%%%%%%%%%%%%%%%%%%%%%%%%%%%%%%%%%%%%%%%%%%%%%%%%%%%%%%%%%%%%%%%%%%%%%%%%%%%%%%%%%%%%%%%%%%
% TMM Derivation
%%%%%%%%%%%%%%%%%%%%%%%%%%%%%%%%%%%%%%%%%%%%%%%%%%%%%%%%%%%%%%%%%%%%%%%%%%%%%%%%%%%%%%%%%%%%%%%%%%%%%%%%%%%%%%%%%%%%%%%%%%%%

\section{Coupling external gauge flux}

The implementation of a uniform synthetic magnetic field for photons and coupling an external gauge field to the edge states hinges on the use of link resonators to couple the site resonators. By vertically displacing the link resonator we can introduce a direction dependent, fixed, hopping phase for the photons hopping between site resonators \cite{Hafezi2011,Hafezi2013}. The fabrication of heaters above the link resonator waveguides enables us to tune this hopping phase. Here, we analyze a system of two site resonators coupled by a link resonator, with a heater on the link resonator waveguide. The heater introduces a phase $2\theta$. We show this system is equivalent to two site rings coupled by an effective coupling rate $J$ and a hopping phase $\theta_{T}$, where $\theta_{T}$ includes the fixed phase incurred due vertical shift of the link resonator and the tunable external gauge flux introduced by the heater.

\linespread{1.0}
\begin{figure}
 \centering
 \includegraphics[width=1.0\textwidth]{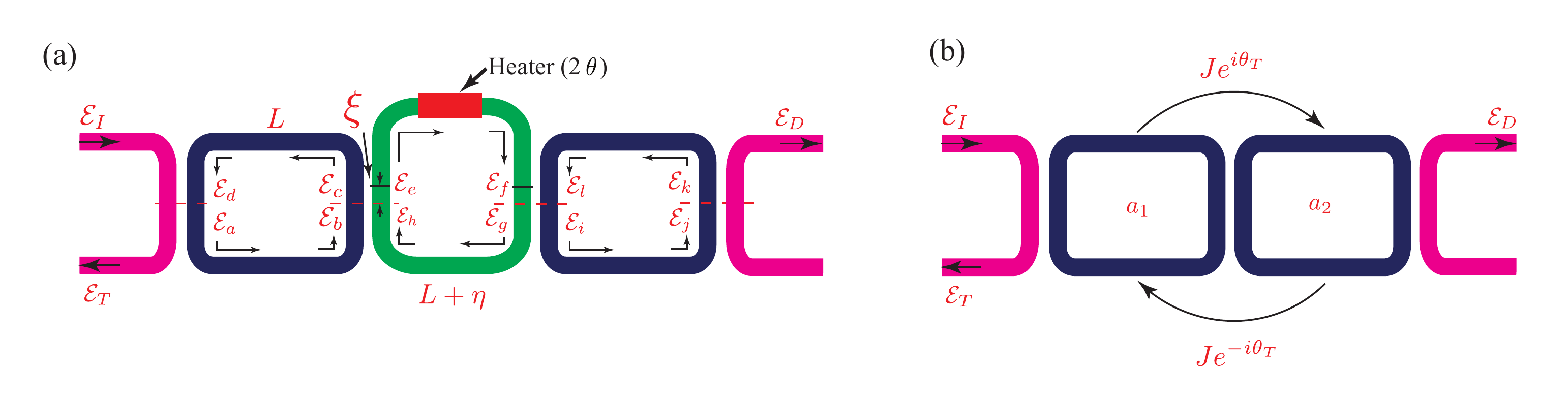}
 \caption{(a) A system of two site resonators, coupled by a link resonator with a heater, and also coupled to input and output waveguides. The figure labels the fields for the transfer matrix analysis. (b) Effective description of the system in coupled mode theory. The effect of the link resonator is contained in the coupling rate $J$ and hopping phase $\theta_{T}$ which accounts for the vertical shift of the link ring and the phase introduced by the heater.}
 \label{fig:S2}
\end{figure}
\linespread{2.0}

We first analyze the system of site rings coupled by a link ring, using the rigorous transfer matrix method. The site resonators have a length $L$ and the link ring $L + \eta$. Furthermore, the link ring is vertically displaced by length $\xi$ and has a heater on the upper arm which introduces a phase $2 \theta$. We label the fields as shown in Fig.~\ref{fig:S2}. Using transfer matrix formalism, we can relate the different field amplitudes as \cite{Hafezi2013}
\bea
\left(\begin{array}{c} \EE_{d}  \\ \EE_{a}  \\ \end{array} \right) &=&
\frac{1}{i\kappa_{i}} \left(
  \begin{array}{cc}
    -t_{i} & 1 \\
    -1 & t_{i} \\
  \end{array}
\right)
\left(\begin{array}{c} \EE_{I} \\ \EE_{\text{T}}  \\ \end{array} \right), \\ % End Equation 1
\left(\begin{array}{c} \EE_{c}  \\ \EE_{b}  \\ \end{array} \right) &=&
  \left(
  \begin{array}{cc}
    e^{-\left(i\beta-\alpha\right) \frac{L}{2}} & 0 \\
    0 & e^{\left(i\beta-\alpha\right) \frac{L}{2}} \\
  \end{array}
  \right)
\left(\begin{array}{c} \EE_{d} \\ \EE_{a}  \\ \end{array} \right), \\ % End Equation 2
\left(\begin{array}{c} \EE_{e}  \\ \EE_{h}  \\ \end{array} \right) &=&
\frac{1}{i\kappa} \left(
  \begin{array}{cc}
    -t & 1 \\
    -1 & t \\
  \end{array}
\right)
\left(\begin{array}{c} \EE_{c} \\ \EE_{b}  \\ \end{array} \right), \\ % End Equation 3
\left(\begin{array}{c} \EE_{f}  \\ \EE_{g}  \\ \end{array} \right) &=&
  \left(
  \begin{array}{cc}
    e^{\left(i\beta-\alpha\right) \left( \frac{L+\eta}{2} + 2 \xi \right)} e^{i 2 \theta} & 0 \\
    0 & e^{-\left(i\beta-\alpha\right) \left( \frac{L+\eta}{2} - 2 \xi \right)} \\
  \end{array}
  \right)
\left(\begin{array}{c} \EE_{e} \\ \EE_{h}  \\ \end{array} \right), \\ % End Equation 4
\left(\begin{array}{c} \EE_{l}  \\ \EE_{i}  \\ \end{array} \right) &=&
\frac{1}{i\kappa} \left(
  \begin{array}{cc}
    -t & 1 \\
    -1 & t \\
  \end{array}
\right)
\left(\begin{array}{c} \EE_{f} \\ \EE_{g}  \\ \end{array} \right), \\ % End Equation 5
\left(\begin{array}{c} \EE_{k}  \\ \EE_{j}  \\ \end{array} \right) &=&
  \left(
  \begin{array}{cc}
    e^{-\left(i\beta-\alpha\right) \frac{L}{2}} & 0 \\
    0 & e^{\left(i\beta-\alpha\right) \frac{L}{2}} \\
  \end{array}
  \right)
\left(\begin{array}{c} \EE_{l} \\ \EE_{i}  \\ \end{array} \right), \\ % End Equation 6
\left(\begin{array}{c} \EE_{D}  \\ 0  \\ \end{array} \right) &=&
\frac{1}{i\kappa_{i}} \left(
  \begin{array}{cc}
    -t_{i} & 1 \\
    -1 & t_{i} \\
  \end{array}
\right)
\left(\begin{array}{c} \EE_{k} \\ \EE_{j}  \\ \end{array} \right)  % End Equation 7
\eea
Here, $t_{i}$ and $k_{i}$ are the field transmission and cross coupling coefficients for coupling of link rings to the input and output ports, and $t,k$ are those for the coupling between the rings. $\beta$ and $\alpha$ are the propagation constant and the loss coefficient in the ring resonator waveguide, respectively. These equations then give the field at the drop port as
\be
\EE_{D} = \frac{e^{i \theta} e^{i 2 \beta \xi} \kappa^2 \kappa_{i}^2 \EE_{I}} {e^{-\left(i\beta+\alpha \right) L}  e^{-\frac{i\beta\eta + i 2 \theta}{2}} \left( e^{i 2 \theta} e^{i \beta \left( L + \eta \right)} \left(t_{i} - t \right)  \left(e^{\alpha L} t - e^{i \beta L} t_{i} \right) + \left(1 - t t_{i} \right) \left( e^{2 \alpha L} - e^{\left(i \beta + \alpha \right) L} t t_{i} \right) \right) }.
\ee
Here we have assumed that the extra length $\eta$ of the link ring and its vertical shift $\xi$ are negligible compared to $L$, so that the loss incurred in these extra lengths is insignificant. In our experimental system, $\eta = 320$ nm, $\xi = 80$ nm and $L \approx 70 \mu$m, which justifies the above assumption.

In the weak coupling limit, i.e., when $\kappa_{i},\kappa \ll 1$ and $t_{i},t \approx 1$, the expression for the drop field can be simplified to
\be
\EE_{D} = \frac{e^{i \theta} e^{i 2 \beta\xi} \kappa^2 \kappa_{io}^2 \EE_{I}}{2 k^2 \left(\frac{\kappa_{i}^2}{2} + \alpha L + i \beta L \right) \cos{\left(\frac{\beta\eta + 2 \theta}{2} \right)} - 2 i \left( \frac{\kappa^{4}}{4} + \left(\frac{\kappa_{i}^{2}}{2} + \alpha L + i \beta L \right)^2 + \right)\sin{\left( \frac{\beta\eta + 2 \theta}{2} \right)}}.
\ee

This system of two site rings coupled by a link ring can be equivalently described by a system involving only the site rings, coupled by coupling rate $J$ and a hopping phase $\pm \theta_{T}$ (Fig.~\ref{fig:S2}b). Specifically, we use single mode approximation for the site rings and employ the coupled mode theory to capture the effect of the link ring in $J$ and $\theta_{T}$. In the coupled mode theory, the rate equations for the time evolution of the site ring energy amplitudes, $a_{1}(t)$ and $a_{2}(t)$, are
\bea
\frac{d a_{1}}{dt} &=& \left(-i\W0-\Kex-\Kin\right) a_{1} - i J e^{-i \theta_{T}} a_{2} -\sqrt{2\Kex} \EE_{\text{I}}\\
\frac{d a_{2}}{dt} &=& \left(-i\W0-\Kex-\Kin\right) a_{2} - i J e^{i \theta_{T}} a_{1},
\eea
where $\Kex$ is the coupling of the site rings to the input and output waveguides and $\Kin$ is the resonator loss rate. A steady-state solution of the above equations for a plane wave excitation of frequency $\Wg$ gives
\bea
a_{1} &=& \frac{\left(i\left(\Wg-\W0\right) - \Kex -\Kin \right) \sqrt{2\Kex} \EE_{I}}{\left(i\left(\Wg-\W0\right) - \Kex -\Kin \right)^2 + J^2}\\
a_{2} &=& \frac{i J e^{i \theta_{T}} a_{1}}{\left(i\left(\Wg-\W0\right) - \Kex -\Kin \right)}.
\eea
Then, the field output at the drop port $\EE_{D}^{\text{CMT}} = \sqrt{2\Kex} a_{2}$ is
\be
\EE_{D}^{\text{CMT}} = \frac{2 i e^{i \theta_{T}} J \Kex  \EE_{I}}{\left(i\left(\Wg-\W0\right) - \Kex -\Kin \right)^2 + J^2}.
\ee

To compare this expression for the drop field to that derived using the transfer matrix approach, we use the relations \cite{Hafezi2013}
\bea
\beta &=& \frac{\left(\W0-\Wg\right)}{\Vg} \\
\Kin &=& \alpha \Vg \\
\Kex &=& \frac{\kappa_{i}^2}{2} \frac{\Vg}{L} \\
J &=& \frac{\kappa^2}{2} \frac{\Vg}{L}.
\eea
The drop field in the transfer matrix formulation is then
\be
\EE_{D} = \frac{2 e^{i\theta_{T}} J \Kex \EE_{I}}{2 J \left(i\left(\W0-\Wg\right)+\Kex+\Kin \right)\cos{(\frac{\beta\eta + 2\theta}{2})} + i \left(J^2 + \left(i\left(\W0-\Wg\right) + \Kex + \Kin \right)^{2} \right) \sin{(\frac{\beta\eta+2\theta}{2})}} ,
\ee
where $\theta_{T} = 2\beta\xi + \theta$. We see that the introduction of a phase $2\theta$ in one of the arms of the link resonator results in an additional hopping phase of $\theta$. Further, for $\beta\eta + 2\theta = \left(\pi, 3\pi, 5\pi, \dots \right)$, i.e.,  when the link ring is anti-resonant to the site rings, the two expressions are identical. Because the link ring is anti-resonant, it does not store any energy and simply acts as a waveguide, without affecting the coupling between the site rings.

When $\beta\eta + 2\theta$ is not an odd-integer multiple of $\pi$, the two expressions for the drop field, derived using the coupled mode theory and the transfer matrix method, are identical if we use an effective coupling rate $J^{\text{eff}} = J/\sin{\left(\frac{\beta\eta+2\theta}{2} \right)}$ and shift the resonance frequency of the site rings as $\W0^{\text{eff}} = \W0 + J \cot{\left(\frac{\beta\eta+2\theta}{2}\right)}$. In our experiment, we design the link rings such that $\beta\eta = \pi$. Further, we distribute the external gauge flux $\theta$ over 12 heaters so that the introduction of external gauge flux $\theta$ results in negligible correction to coupling rate $J$ and a small shift in the resonance frequency of the site rings. For example, for $\theta = 2\pi$ divided over 12 link rings, the correction to the coupling rate $J$ is only. 3.5$\%$. However, the shift in resonance frequency $\Delta\W0$ is $0.27 J$, which adds to disorder in the lattice.

Our system supports two degenerate spin states - circulating CW and CCW in the ring resonators. The above discussion is for CCW state. Following similar procedure for CW state (spin flipped) results in a negative sign for the gauge flux theta. \black Also note that coupling a gauge flux by locally heating an arm of the link resonator is not the same as increasing the length of the resonator. In particular, the gauge flux $\theta$ introduced here has opposite sign for opposite direction of hopping (right or left, see Fig.\ref{fig:S2} b). \black

%%%%%%%%%%%%%%%%%%%%%%%%%%%%%%%%%%%%%%%%%%%%%%%%%%%%%%%%%%%%%%%%%%%%%%%%%%%%%%%%%%%%%%%%%%%%%%%%%%%%%%%%%%%%%%%%%%%%%%%%%%%%%
% Eigenvalues of Square Annulus
%%%%%%%%%%%%%%%%%%%%%%%%%%%%%%%%%%%%%%%%%%%%%%%%%%%%%%%%%%%%%%%%%%%%%%%%%%%%%%%%%%%%%%%%%%%%%%%%%%%%%%%%%%%%%%%%%%%%%%%%%%%%

\section{Eigenvalues of square annulus as a function of external flux}

\linespread{1.0}
\begin{figure}
 \centering
 \includegraphics[width=0.8\textwidth]{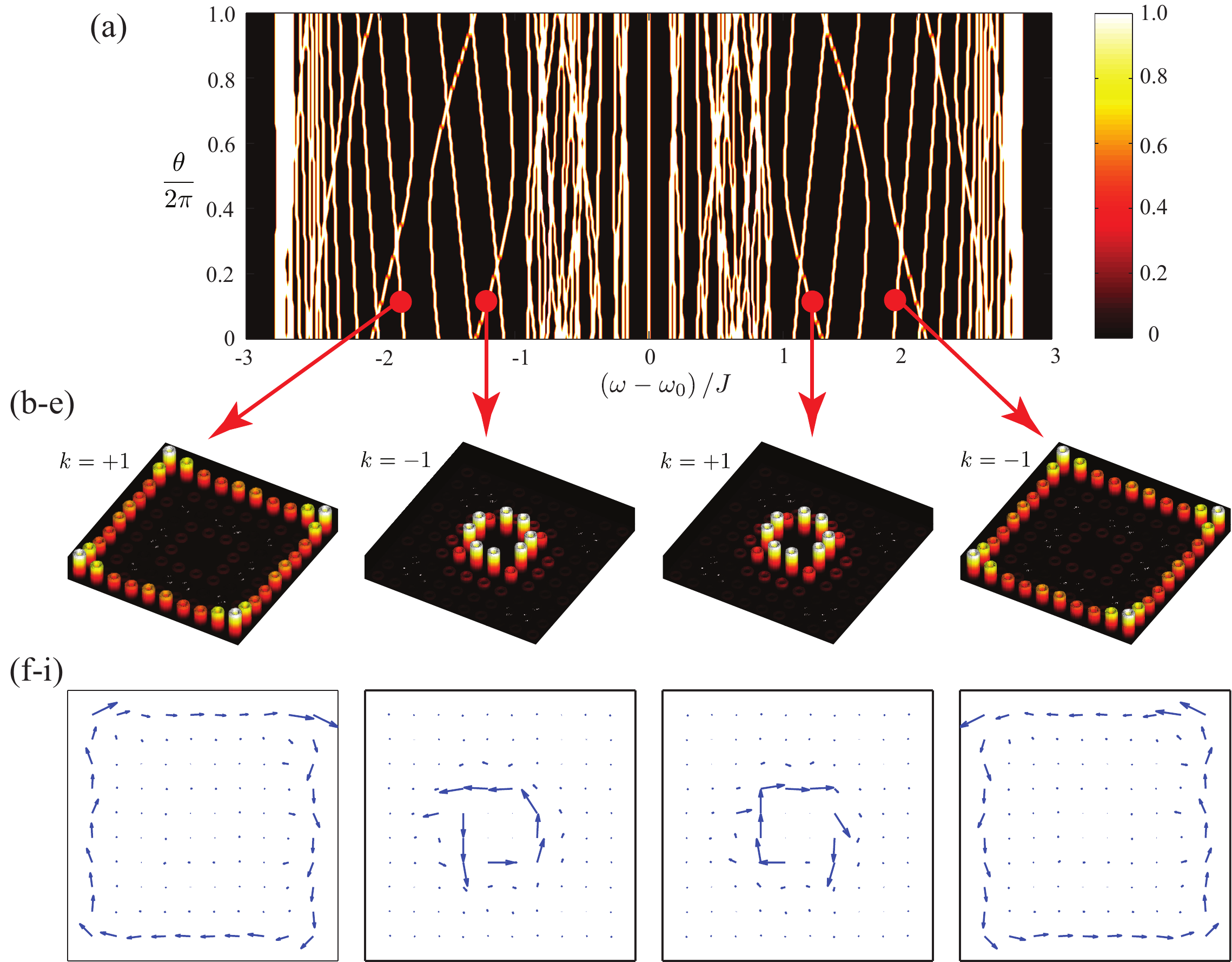}
 \caption[Shifting of eigenvalues in a square lattice.]{(a) Shifting of eigenvalues as a function of external magnetic flux $\theta$. The annulus now supports two sets of edge states, at the outer (b,e) and the inner edges (c,d). In a given bandgap, the outer (f,i) and the inner (g,h) edge states circulate around the lattice in opposite directions because of their opposite group velocity. The arrows indicate direction of group velocity (energy flow). Furthermore, in the energy spectrum (a), the clockwise and the counterclockwise circulating states move in opposite directions because of their winding number which is $\pm$ 1. }
 \label{fig:S1}
\end{figure}
\linespread{2.0}

The 2D annulus geometry of site resonators (shown in Fig. 1) can be described using coupled-mode theory with a Hamiltonian \cite{Hafezi2011,Hafezi2013} given as
\begin{align}
  H_{0} = & - J \sum_{x,y} \hat{a}_{x+1,y}^{\dag} \hat{a}_{x,y}^{} e^{-i y \phi_{0}} e^{\pm i \phi_{t} \delta_{x,y;edge}} + \hat{a}_{x,y}^{\dag} \hat{a}_{x+1,y}^{} e^{i y \phi_{0}} e^{\mp i \phi_{t} \delta_{x,y;edge}} \\
   & + \hat{a}_{x,y+1}^{\dag} \hat{a}_{x,y}^{} e^{\pm i \phi_{t} \delta_{x,y;edge}} + \hat{a}_{x,y}^{\dag} \hat{a}_{x,y+1}^{} e^{\mp i \phi_{t} \delta_{x,y;edge}},
\end{align}
where $\hat{a}_{x,y}$ and $\hat{a}^{\dag}_{x,y}$ are the photon annihilation and creation operators, respectively, at a site resonator with position index $x,y$, and $J$ is the hopping rate for photons. The hopping phase $\phi_{0}$ along the $x-$axis gives rise to a uniform magnetic field in the bulk with flux $\phi_{0}$ per plaquette \cite{Hafezi2011, Hafezi2013}. In addition to this uniform bulk field, we couple a tunable gauge flux $\theta$ only to the outer (or the inner) edge of the system (Fig. 1). The tunable hopping phase $\phi_{t} = \frac{\theta}{N}$, where $N = 12$ is the number of heaters used on the edge to achieve an accumulated  flux $\theta$. The sign of the tunable phase term is $-ve$ for top and left edges of the annulus; and $+ve$ for bottom and right edges of the annulus.\\

Fig.~\ref{fig:S1} shows the eigenvalues of this annulus as a function of the external flux $\theta$. This system supports two sets of CW and CCW propagating edge states, one each on the outer and the inner edges. Furthermore, edge states on the outer edge (e.g. CW) are paired with edge states of opposite chirality (CW) on the inner edge, i.e.,  they occupy the same bandgap (left). To illustrate the spectral flow of these edge states as a function of external gauge flux, we couple the gauge flux to both the outer and the inner edges of the system. We observe that with an increase in the external flux, all the edge states shift outwards or inwards. The CW-propagating outer edge states in the first bandgap move toward lower energies, whereas the CCW-propagating outer edge states in the third bandgap move toward higher energies. The inner edge states move exactly opposite to the outer edge states. More importantly, for a $2\pi$ increase in magnetic flux, each set of states shift exactly by one, their winding number. The direction of shift indicates the sign of the winding number, as is shown in Fig.~\ref{fig:S1}. After a $2\pi$ increase in the flux, the spectrum returns back to its original shape with $\theta = 0$. Note that in the measured transmission data shown in the main text, we observe transmission resonances corresponding only to the outer edge states. This is because in our system, the input and output waveguide couplers are coupled only to the outer edge and their coupling to inner edge is exponentially suppressed (Fig. 1). \black Furthermore, it is instructive to compare the flow of resonances observed here to that in a Fabry-Perot cavity. As we discussed, introduction of an additional gauge flux is not the same as increasing the length of the resonator. In particular, the sign of the gauge flux is opposite for CW and CCW propagating states, and therefore, these resonances move in opposite directions. In contrast, increasing the length of the Fabry-Perot cavity shifts all resonances towards lower frequencies.

Also note that in the transmission spectra presented in the main text (Fig. 2, 3), only the outer edge contributes to the transmission spectra. This is because the coupling between the probe waveguides to the inner edge states is exponentially suppressed as $e^{-d/l_{0}}$, where $d$ is the number of rings between the outer and the inner edge (in this case 4) and $l_{0} = 1/\sqrt{\phi_{0}}\approx 0.8$ is the magnetic length in units of the lattice spacing.

%%%%%%%%%%%%%%%%%%%%%%%%%%%%%%%%%%%%%%%%%%%%%%%%%%%%%%%%%%%%%%%%%%%%%%%%%%%%%%%%%%%%%%%%%%%%%%%%%%%%%%%%%%%%%%%%%%%%%%%%%%%%
% Heater Calibration
%%%%%%%%%%%%%%%%%%%%%%%%%%%%%%%%%%%%%%%%%%%%%%%%%%%%%%%%%%%%%%%%%%%%%%%%%%%%%%%%%%%%%%%%%%%%%%%%%%%%%%%%%%%%%%%%%%%%%%%%%%%%

\section{Heater calibration}

\linespread{1.0}
\begin{figure}
 \centering
 \includegraphics[width=0.8\textwidth]{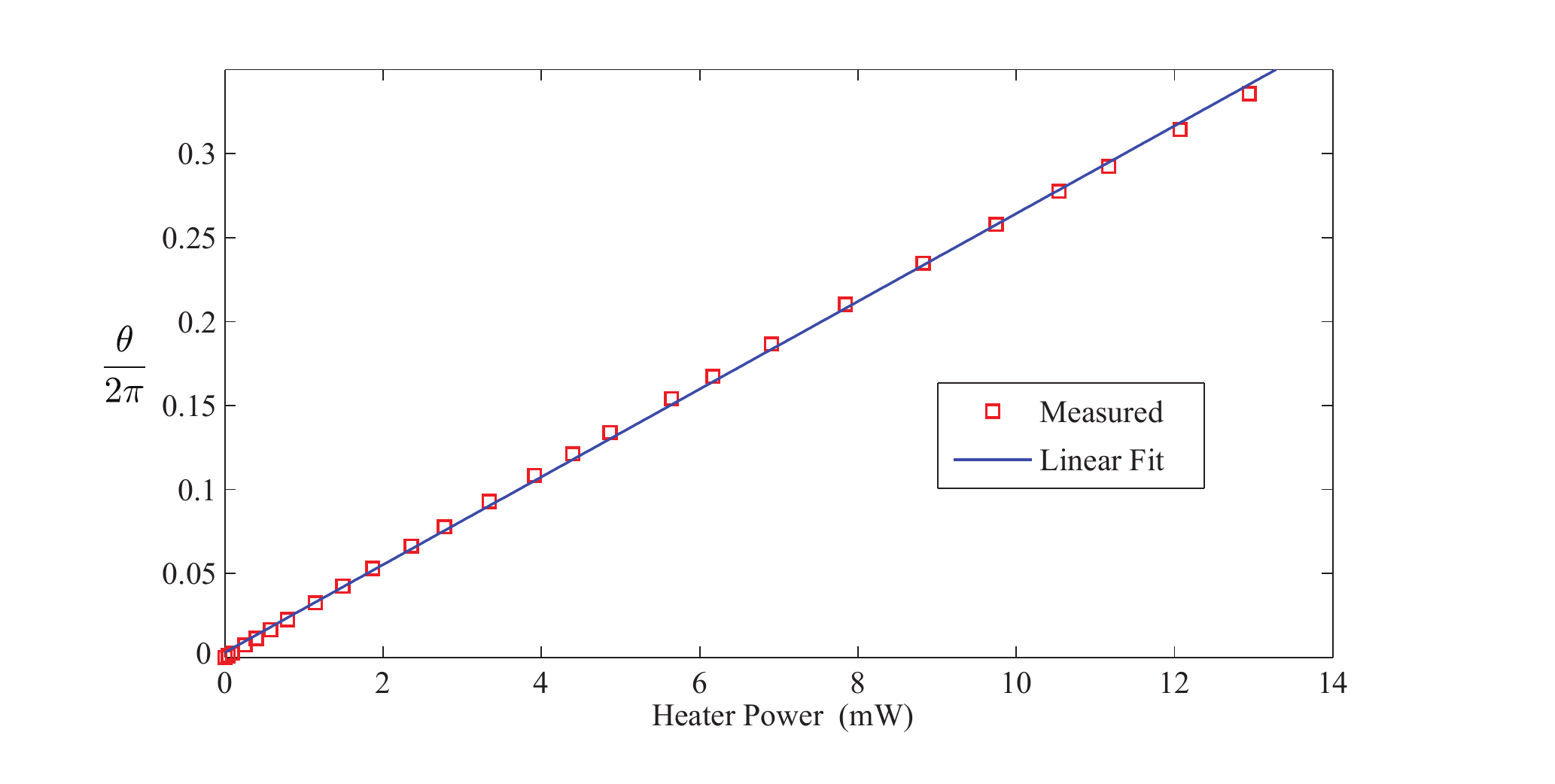}
 \caption{Measured phase shift in an ADF, as a function of the heater power. The acquired phase shift (in units of 2$\pi$) increases linearly with the heater power (in mW), with a slope of 0.026(1).}
 \label{fig:S3}
\end{figure}
\linespread{2.0}

We used an add-drop filter (ADF - a ring coupled to two waveguides) with a heater, to calibrate the phase shift acquired for a unit heater power. The ADF ring has exactly the same dimensions as the rings in the 2D array. The ring waveguides are designed to be 510 nm wide to ensure single mode (TE) operation at telecom wavelengths and the typical ring perimeter is $\approx 70 \mu$m. The coupling gap between the ring waveguides is 140 nm which gives a coupling rate $J = 40.6$ GHz. The heater introduces a phase shift $\theta$ in the ADF ring waveguide and shifts its resonance frequency by $\Delta\W0$, such that
\begin{equation}
\theta = \frac{\Delta\W0}{FSR} 2\pi,
\end{equation}
where $FSR = 2\pi \frac{v_{g}}{L}$ is the free spectral range of the ring, $v_{g}$ is the group velocity and $L$ is the length of the ring. Measuring the resonance frequency shift as a function of heater power gives us an estimate of the phase difference introduced by the heater. We used an ADF ring which had exactly the same dimensions as the rings in the square annulus and the chain of resonators. Fig.~\ref{fig:S3} shows the observed phase shift as a function of heater power. The heater resistance was $\approx$ 120 Ohms and the voltage applied to the heater ranged from 0-3 Volts. We see that the acquired phase increases linearly with the heater power.

For the square annulus and the chain geometry, we used this calibration curve to estimate $\theta$ for a given voltage applied to the heaters. For the annulus, all the 12 heaters in the outer (and the inner) loop (Fig.~1c)) were serially connected. The measured resistance was 1.4 KOhms and the applied heater voltage across the loop ranged from $\approx$ 8-14 Volts, which introduced the phase $\theta = \pi- 3\pi$. Similarly, in the chain of rings (Fig.~4a), all the 20 heaters were connected serially, giving a total resistance of 2.31 KOhms. The total voltage applied across the heaters ranged from $\approx 0-13$ Volts, for $\theta = 0 - 2\pi$. Furthermore, as we saw in the previous section, the transfer matrix analysis of a system of two site rings, coupled by a link ring, shows that if the heater introduces an extra phase $2 \theta$ in one of the arms, the effective hopping phase (using CMT) between the link rings is $\theta$. We have included this factor of 2 in our calculations.

\section{Disorder estimation}

In our design, the heaters are placed 600 nm above the link resonator waveguides. Because of this, the heat is not confined only to the ring beneath the heater but instead also heats the neighboring site resonators. Using a three ring device where two site rings are coupled by a link ring, we estimate that for every unit shift in resonance frequency of the link ring, the site rings move by 0.3 units. The small curvature observed in the shifting of edge states is also a result of this disorder. We have accounted for this disorder in our simulation. Moreover, because of this disorder we find that for the square annulus, the transmission spectrum retains its shape in the range $\theta \approx \pi - 3\pi$. In our measurements reported in Fig.~2, the external gauge flux has been normalized to $\theta = 0 - 2\pi$.

Furthermore, because of this nonlocal behavior of the heaters, the device spectrum shifts towards lower frequencies as we increase the voltage applied to the heaters. We have adjusted for this overall shift of the transmission spectra. For each spectrum, we choose two edge state peaks, one each on CW and CCW bands and align all the spectra to the center of these two peaks. The resulting plot thus shows only the differential shift between the two edge state bands. For the ring geometry, the measured spectrum at $\theta=2\pi$ in Fig. 4d has been scaled along the frequency axis (by a factor of 0.98), to offset the dispersion effect (due to the overall shift) which results in slight broadening of the spectrum.

%%%%%%%%%%%%%%%%%%%%%%%%%%%%%%%%%%%%%%%%%%%%%%%%%%%%%%%%%%%%%%%%%%%%%%%%%%%%%%%%%%%%%%%%%%%%%%%%%%%%%%%%%%%%%%%%%%%%%%%%%%%%
% Comparison with Laughlin's Charge Pump and Hall drift measurements
%%%%%%%%%%%%%%%%%%%%%%%%%%%%%%%%%%%%%%%%%%%%%%%%%%%%%%%%%%%%%%%%%%%%%%%%%%%%%%%%%%%%%%%%%%%%%%%%%%%%%%%%%%%%%%%%%%%%%%%%%%%%

\section{Comparison with Laughlin's charge pump}

The spectral flow observed here is very similar to Laughlin's charge pump \cite{Laughlin1981, Halperin1982}. The Laughlin's pump was experimentally observed for 1D dimensional photonic systems \cite{Kraus2012, Hu2014}, and proposed as a way to measure winding number in 2D photonic systems \cite{Hafezi2014}. However, our scheme requires manipulation and probe only at the boundary of the system, unlike Laughlin's case where the gauge is coupled to both the edge and the bulk. Therefore, our scheme avoids technical complications that may arise due to the bulk manipulation which is required in Ref.\cite{Hafezi2014}.

%%%%%%%%%%%%%%%%%%%%%%%%%%%%%%%%%%%%%%%%%%%%%%%%%%%%%%%%%%%%%%%%%%%%%%%%%%%%%%%%%%%%%%%%%%%%%%%%%%%%%%%%%%%%%%%%%%%%%%%%%%%%
% Bibliography
%%%%%%%%%%%%%%%%%%%%%%%%%%%%%%%%%%%%%%%%%%%%%%%%%%%%%%%%%%%%%%%%%%%%%%%%%%%%%%%%%%%%%%%%%%%%%%%%%%%%%%%%%%%%%%%%%%%%%%%%%%%%
\bibliographystyle{NatureMag}
\bibliography{Winding_Biblio}